\documentstyle[epsf]{mn2e}

\title[The X-ray Binary Analogy to the First AGN QPO]
{The X-ray Binary Analogy to the First AGN QPO}

\author[M.Middleton, C. Done]
{Matthew Middleton$^1$, Chris Done$^1$\\
$^1$Department of Physics, University of Durham, South Road, Durham
DH1 3LE,
UK\\
}

\pagerange{\pageref{firstpage}--\pageref{lastpage}} \pubyear{2009}

\begin{document}

\topmargin = -0.5cm

\maketitle

\label{firstpage}

\begin{abstract}

The Narrow Line Seyfert 1 galaxy REJ1034+396 is so far unique amongst
AGN in showing a Quasi-periodic oscillation (QPO) in its variability
power spectrum. There are multiple types of QPO seen in black hole
binary (BHB) systems, so we need to identify which BHB QPO corresponds
to the one seen in the AGN.  A key hint is the `hot disc dominated'
energy spectrum of REJ1034+396 which is sufficiently unusual that it
suggests a mildly super-Eddington flow, also favoured by the most
recent mass estimates for the AGN. This suggests the 67Hz QPO seen
occasionally in the mildly super-Eddington BHB GRS 1915+105 as the
most likely counterpart, assuming mass scaling of the QPO
frequency. This is supported by the fact that these data from
GRS 1915+105 have an energy spectrum which is also dominated by a `hot
disc' component. Here we show that the underlying broad band power
spectral shape and normalisation are also similar, providing further
consistency checks for this identification. Thus the AGN QPO adds to
the growing evidence for a simple mass scaling of the accretion flow
properties between stellar and supermassive black holes.

\end{abstract}
\begin{keywords}  accretion, accretion discs -- X-rays: binaries, black hole
\end{keywords}

\section{Introduction}

Active galaxies (AGN) and accreting Black hole binaries (BHB) are the
most energetic and dynamic systems in the Universe, with intrinsic
variability seen over a broad range of timescales. This variability is
most often quantified via a power spectrum, showing the (squared)
amplitude of variability as a function of frequency. For BHB this
spectrum can be very approximately described as band limited noise,
with a `flat top' in $\nu P(\nu )$ (equal variability power per decade
in frequency) i.e. $P(\nu)\propto\nu^{-1}$. This extends between a low
and high frequency break, $\nu_{b}$, below which the PDS is
$P(\nu)\propto \nu^0$, and $\nu_{h}$, above which the spectrum
steepens to $P(\nu)\propto \nu^{-2}$. While this is adequate to describe
the limited statistics of AGN power spectra (see the compilation of
Markowicz et al. 2003), the excellent data from BHB show that band
limited noise is intrinsically bumpy, and is better represented by a
series (generally 4--5) of peaked noise components (Belloni \&
Hasinger 1990; Psaltis, Belloni \& van der Klis 1999; Nowak 2000;
Belloni, Psaltis \& van der Klis 2002). Superimposed on this
continuum are a series of narrower features, referred to as
Quasi-periodic oscillations (QPOs).  These QPOs come in two distinct
categories, termed low and high frequency QPOs respectively, and
their properties depend strongly on mass accretion rate (see e.g. the
reviews by van der Klis 2004; Remillard \& McClintock 2006; Done,
Gierli{\'n}ski \& Kubota 2007, hereafter DGK07).

\begin{figure}
\begin{center}
\begin{tabular}{l}
 \epsfxsize=8cm \epsfbox{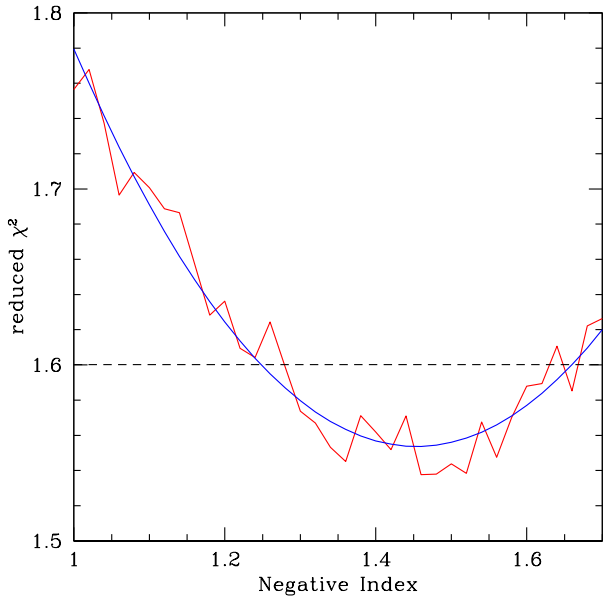}
\end{tabular}
\end{center}
\caption{Reduced $\chi^{2}$ (58 d.o.f.) plot for single power-law
power spectrum simulations for the `in phase' lightcurve. We fit this
to a quadratic function to smooth over the intrinsic stochastic
effects. We use this to define the best fit index and its uncertainty
($\Delta \chi^{2}=2.7$ i.e. 90 \% confidence as shown by the dashed
horizontal line) as -1.45$^{+0.20}_{-0.21}$. This is consistent with
the analytical result of $-1.35\pm 0.18$ (Gierli{\'n}ski et
al. 2008).}
\label{fig:l}
\end{figure}

There is now considerable evidence to suggest that AGN and BHB are
mass-scaled analogies and share the same underlying physics of
accretion, notably the Radio/X-ray fundamental plane relations
(Merloni, Heinz \& di Matteo 2003; Falcke, K{\"o}rding \& Markoff
2004), variability scaling (McHardy et al. 2006; McHardy 2009) and
mass-accretion rate dependant spectral behaviour (Pounds, Done \&
Osborne 1995; Middleton, Done \& Schurch 2008; Middleton, Done \&
Gierli{\'n}ski 2007). If this is the case then we would naturally
expect QPOs to be visible in the power spectra of AGN at frequencies
inversely scaled with the mass. However, these much longer timescales
make detecting such QPOs in AGN very challenging (Vaughan \& Uttley
2006). Many claims for such periodic signals have been made, but these
all failed to properly determine the significance of the QPO on top of
the (uncertain determination) of the underlying red noise power
spectrum (see the discussion in Vaughan 2005).  By contrast, our
discovery of a $\sim$ 1 hour periodic signal from a Narrow Line
Seyfert 1 (NLS1), REJ1034+396 is highly significant ($>$99\%)
\textit{with} correct modelling of the red noise spectrum
(Gierli{\'n}ski et al. 2008). This gives the first detected AGN QPO to
compare with the BHB.

\section{Mass scaling}

Beyond simply detecting the QPO, we need to identify it with one of
the multiple BHB QPOs. Up until recently this has been a problem as
the mass of the AGN black hole was not well constrained, with
different techniques (H$\beta$ line width, stellar velocity
dispersion) giving mass estimates which differ by a factor of 30-50
(Gierli{\'n}ski et al. 2008). However, a recent reanalysis of the SDSS
data from this object shows that both these black hole mass tracers
can give consistent estimates for the black hole mass with careful
decomposition of the different spectral components, giving
1-4$\times$10$^{6}$M$_{\odot}$ (Bian \& Huang 2009).  This predicts an
analogous signal from a 10M$_{\odot}$ BHB between 27 and 108~Hz if
a simple linear mass scaling holds for this QPO. 

\begin{figure}
\begin{center}
\begin{tabular}{l}
 \epsfxsize=8cm \epsfbox{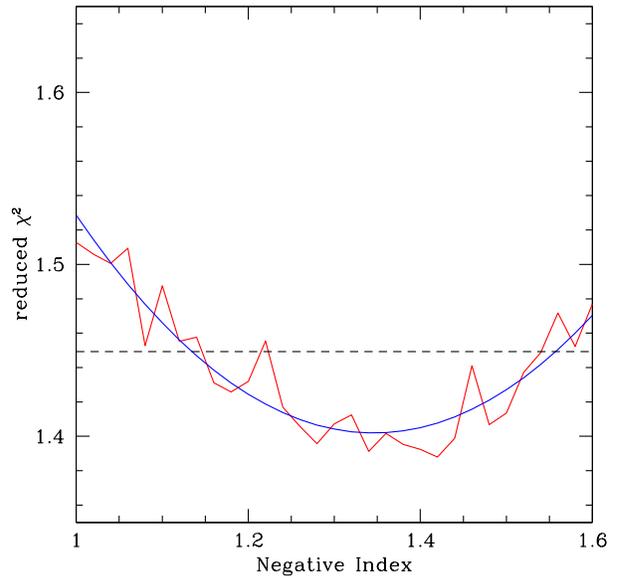}
\end{tabular}
\end{center}
\caption{As for Fig 1. but excluding the single frequency bin of the QPO
in the `in phase' segment (57 d.o.f.). 
The result is consistent with that of simulations including
the QPO over the same segment, as expected as the QPO is fairly 
central in the power spectrum so this power does not distort 
the index.}
\label{fig:l}
\end{figure}

Placing this mass estimate together with the bolometric luminosity of
$\sim 5\times 10^{44}$ergs s$^{-1}$ implies 1$<$L/L$_{Edd}$$<$4. This is
supported by the distinctive nature of the optical/UV/X-ray SED of
REJ1034+396, where the bolometric luminosity is dominated by a very
luminous component peaking in the EUV (Puchnarewicz et al. 2001;
Middleton, Done \& Gierli{\'n}ski. 2007). This cannot be fit by
standard disc models, but can be approximated by the super-Eddington
slim disc models (Mineshige et al. 2000; Wang \& Netzer 2003, Haba et
al. 2008).

The type and characteristic frequency of QPOs seen from BHB depend on
mass accretion rate as well as BHB mass. Thus we need to study a
super-Eddington BHB in order to find an analogue with the AGN
QPO. GRS~1915+105 is the only BHB which spends significant time at
(probably) super-Eddington accretion rates (Done, Wardzinski \&
Gierli{\'n}ski 2004). This has a mass of $\sim$14M$_{\odot}$ (Greiner,
Cuby \& McCaughrean 2001; Harlaftis \& Greiner 2004), so a simple mass
scaling of the AGN QPO predicts the counterpart is between 19 and
77Hz. GRS1915+105 does indeed occasionally show a QPO at 67~Hz (with a
probable harmonic at 41~Hz), making this the most likely candidate.

This means that the data are consistent with the simplest model, which
is that the QPO seen at a given luminosity/spectral state in BHB
scales linearly with mass up to AGN. However, this only makes use of
the information on the QPO frequency. This QPO sits on top of an
underlying broad band continuum noise power spectrum, and the shape
and normalisation of this broadband noise is likewise dependent on
luminosity/spectral state in BHB. Here we use the additional
information from this noise power to check the QPO identification.

\section{Power spectral phenomenology in BHB}

\subsection{Low-frequency QPOs} 

These are subdivided into 3 classes, types A, B and C (see
e.g. Casella et al. 2005), not to be confused with the spectral states
in GRS~1915+105 which are given the same name (Belloni et al. 1997)!

The C class LFQPO is almost always present in the low/hard state of
BHB. It has frequency which {\em increases} between $\sim$ 0.01-6~Hz
as the spectrum softens from a dim, harder, low/hard state to a
bright, softer low/hard state, through to the hard intermediate state
(Belloni et al. 2005). There is a correlated increase in strength,
coherence and harmonic structure of the QPO (van der Klis 2004;
Remillard \& McClintock 2006). Its spectrum closely follows that of
the harder X-ray Comptonisation component rather than the soft disc
emission (Rodriguez et al. 2002; Sobolewska \& Zycki 2003; 2006), so
it is most probably produced by some fundamental mode of the corona,
not of the disc. 

The changes in underlying power spectral shape are correlated with the
energy spectral changes and the QPO frequency changes.  Very
approximately, the type C QPO frequency increases together with the
low frequency break, $\nu_{b}$, (Psaltis et al. 1999), while the high
frequency break remains constant. Thus the broad band power spectrum
progressively narrows, losing the low frequency power (Gierli{\'n}ski,
Niko{\l}ajuk \& Czerny 2008). The normalisation of the `flat top' remains
roughly constant at $\nu P(\nu)\approx 0.01$ when the power spectrum
is normalised to the fractional fluctuation power $(\sigma/I)^2$ at
high energies. The bandpass is important as the disc component becomes
progressively stronger as the spectrum softens. The disc remains
mostly constant on the short timescales of the power spectrum, so
dilutes the variability of the higher energy tail (Churazov et al.
2001; Done \& Gierli{\'n}ski 2005).

In the context of the truncated disc/hot inner flow models for the
low/hard state, all these observations can be interpreted if the inner
edge of the thin disc decreases as a function of mass accretion
rate. The increasing overlap between the disc and hot flow means more
seed photons from the disc intercept the hot flow, giving stronger
Compton cooling and softer spectra.  The moving inner radius means
that the section of the hot flow which is unconstrained by the disc
gets smaller, so increasing all its characteristic variability
frequencies, while the smaller radial extent means the QPO can become
stronger and more coherent (DKG07). This qualitative picture can be
made quantitative, using vertical (Lense-Thirring) precession of the
hot flow to make the QPO (Ingram, Done \& Fragile 2009), and
turbulence in the hot flow to make the broadband power (Ingram \& Done
2009), but such models are still at the speculative stage.

The type B LFQPO is seen as the source continues to soften after the
highest type C LFQPO from the hard intermediate state to the soft
intermediate state. This is a rather subtle change in spectrum, but
triggers an abrupt collapse of the underlying broad band noise power
spectrum, leaving the QPO (and its multiple harmonics) dramatically
enhanced in prominence (Remillard et al. 2002; Belloni et al. 2005).

As the spectrum continues to soften in the soft intermediate state,
class A LFQPOs can be seen. These are weaker and much broader (so
there is no harmonic structure discernible) than the type B LFQPOs, but
appear at similar frequencies, and again are seen against a very low
continuum noise power (Remillard et al. 2002). Then the QPOs disappear
altogether as the spectrum continues to soften into a completely disc
dominated state (Remillard \& McClintock 2006; Casella et al. 2005).

While the type C LFQPO is ubiquitous in the low/hard and hard
intermediate states, types B and A are much rarer, seen only in the 
soft intermediate state. 

\begin{figure}
\begin{center}
\begin{tabular}{l}
 \epsfxsize=8cm \epsfbox{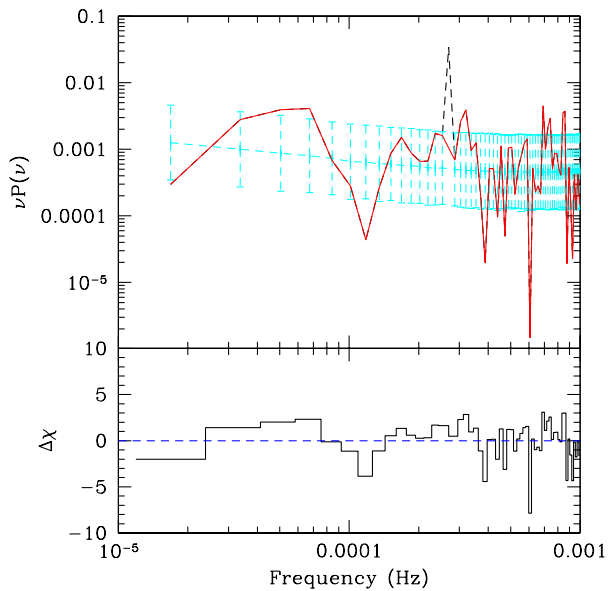}
\end{tabular}
\end{center}
\caption{The upper panel shows the best-fit average simulated power
spectrum (cyan) to that of the `in phase' data (red) without the QPO
(shown as the dashed black line). This 
assumes a single power-law form for the intrinsic 
fluctuations, together with statistical (white) noise. The dispersion in the 
simulations is shown by the dashed error bars (cyan). 
The lower panel shows residuals to this best fit.} 
\label{fig:l}
\end{figure}

\subsection{High-frequency QPOs}

Transient high-frequency QPOs (HFQPOs) are detected in several BHB
sources (GRO J1655-40, XTE J1550-564, GRS~1915+105, 4U 1630-47 and XTE
J1859+226). These appear to be constant in frequency despite changes
in luminosity by a factor of 3-8, so these may be a stable signature
of the accretion system. The values are consistent with being directly
proportional to the mass of the black hole, though there are only 3
data points and the masses only range from $6-14M_\odot$ (Remillard \&
McClintock 2006). They are often seen in 3:2 ratio, implying some sort
of resonance sets the frequencies, but the exact nature of this
resonance is yet to be clearly determined (Abramowicz \& Kluzniak
2001; Kluzniak et al. 2004; T{\"o}r{\"o}k et al. 2005; Kato et
al. 2004; Rezzolla et al. 2003; Fragile et al. 2005).

These seem to be detected only in the soft intermediate state
i.e. are seen simultaneously with type B and A LFQPOs and
not with type C LFQPOs (Remillard et al. 2002). It is extremely
difficult to accurately assess the underlying noise power spectral
shape at such high frequencies as this is dominated by the statistical
white noise, but the co-added spectra of Remillard et al. (2002)
suggest that these high frequency QPOs sit on a broad band continuum
which has a $\nu^{-1}$ shape (most probably another Lorentzian
component) but with a normalisation which is over an order of
magnitude below that seen for the lower frequency `flat top' noise
which accompanies the type C LFQPOs (Remillard et al. 2002). While the
normalisation of the power spectrum strongly depends on the energy
bandpass in BHB due to the different variability of disc and
Comptonisation components (e.g. Churazov et al. 2001; Done \&
Gierli{\'n}ski 2005), the study of Remillard et al. (2002) used only data
above 6~keV where the disc makes little contribution. Hence this
difference in normalisation is probably intrinsic rather than due to
dilution of the variability by the constant disc.

\begin{figure}
\begin{center}
\begin{tabular}{l}
 \epsfxsize=8cm \epsfbox{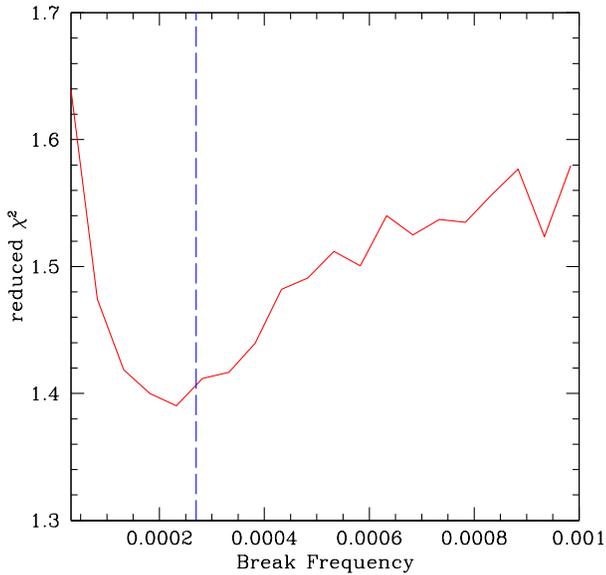}
\end{tabular}
\end{center}
\caption{As in Fig 1 but for an underlying 
broken power-law (indices from -1 to -2) spectral shape 
of the `in phase' segment of the lightcurve
without the QPO. The $\chi^2$ of the best fit of  a break 
close to the QPO frequency (2.7$\times$10$^-4$~Hz: vertical dashed line) is
similar to that for the single
power-law fit in Fig 2, so this is not a significantly better
description of the spectral shape.} \label{fig:l}
\end{figure}

\subsection{Additional QPOs seen only in GRS~1915+105}

GRS~1915+105 is highly luminous, so never gets to the low/hard
state. Instead it ranges between the hard intermediate state, soft
intermediate state and disc dominated states. However, it also shows
spectra which are unlike any of the standard states, being dominated
by low temperature Comptonisation of the disc together with a very
weak hard tail (Zdziarski et al. 2003; Done, Wardzinski \& Gierlinski
2004; Belloni 2009). In this state it can show some particularly
unique QPOs, including very strong, narrow QPOs with complex harmonic
structure at tens of milliHertz (Morgan, Remillard \& Greiner
1997). These are startlingly similar to the QPO frequencies seen in
the Ultra-Luminous X-ray sources M82 X-1 (Strohmayer \& Mushotzky
2003) and NGC5408 X-1 (Strohmayer at al. 2007). While these low
frequencies clearly imply intermediate mass black holes if they are
identified with the standard LFQPO, it is plain from the data of
GRS~1915+105 that such low frequencies can also be formed from
super-Eddington accretion onto a 14~$M_\odot$ black hole (Heil,
Vaughan \& Roberts 2009).  Simultaneously with this very low frequency
QPO, GRS~1915+105 can also show an occasional QPO at 67~Hz.  This has
only been strongly detected in a few datasets (Morgan, Remilllard \&
Greiner 1997) plus a few other weaker detections (Morgan, Remilllard
\& Greiner 1997; Ueda et al. 2009), together with a (potential
harmonic) feature at 41~Hz (McClintock \& Remillard 2006). The
statistical noise is not so dominant as for the HFQPOs as these are at
lower frequencies, so the underlying continuum noise power is somewhat
easier to constrain. This has a complex shape, but the normalisation
underneath the QPO appears to be $\nu P(\nu)\sim 10^{-4}$ from Morgan
et al. (1997). However, this used power spectra across the full energy
bandpass. Concentrating instead on the high energy band gives $\nu
P(\nu)\sim 0.001$ (Remillard \& McClintock 2006).

\begin{figure}
\begin{center}
\begin{tabular}{l}
 \epsfxsize=8cm \epsfbox{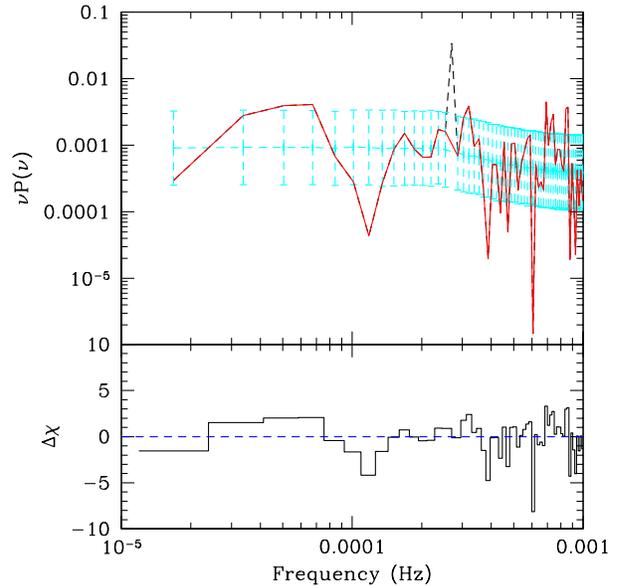}
\end{tabular}
\end{center}
\caption{As in Fig 3 but for a broken power-law (indices from -1 to -2)
fit to the `in phase' power spectrum without the QPO. 
} \label{fig:l}
\end{figure}

\section{Lightcurve Simulations}

The analytical method of fitting the broadband shape of the power
spectrum of Vaughan (2005) is only applicable to a single power-law
continuum noise shape, so does not include intrinsic breaks or the
effects of statistics (white noise at high frequencies). Instead we
simulate an ensemble of lightcurves with given power spectral
properties using the method of Done et al. (1992), using the corrected 
power spectral statistics (Timmer \& Koenig 1995),  as developed by
Uttley et al. (2002). 

The real data are evenly sampled and binned on timescale $\Delta t$
and span a total time of $T_{data}$. This has mean, $I$, total
variance, $\sigma^2_{data}=\sigma^2_{int} +\sigma^2_{err}$, where
$\sigma^2_{int}$ is due to the intrinsic variability of the source and
$\sigma^2_{err}$ is the additional variance added by statistical
uncertainties.  We generate a lightcurve using the fast Fourier
transform method of Timmer \& Koenig (1995), spanning a timescale
N$T_{data}$ (where N$\gg$1) with some intrinsic power spectral shape,
$P(f) sin^2 (2\pi f \Delta t) / 2\pi f \Delta t$, where the $\sin^2$
term accounts for the suppression of high frequency power due to
binning (Done et al.  1992). We snip this up into N separate
lightcurves of length $T_{data}$. For each lightcurve we measure
$\sigma^2_{i,int}$ and $I_i$, and then scale so that the average mean
and variance of the ensemble is the same as that in the real data. We
then add white noise with the same properties as in the real data,
take the power spectra of the simulated lightcurves and use these to
define the statistics of the power spectral distribution.

\begin{figure}
\begin{center}
\begin{tabular}{l}
 \epsfxsize=8cm \epsfbox{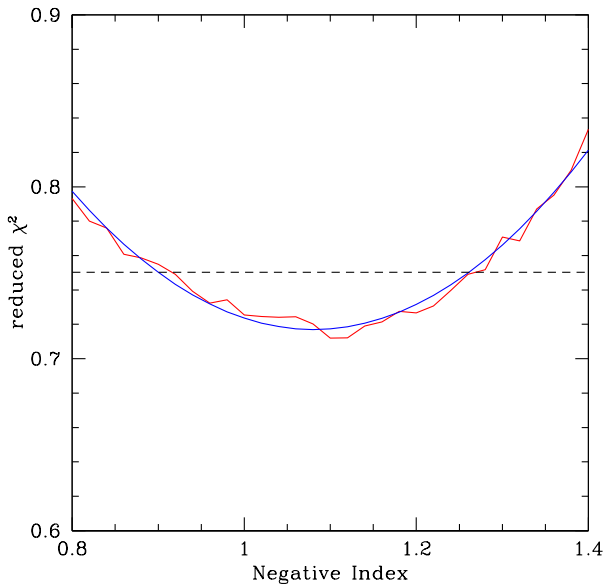}
\end{tabular}
\end{center}
\caption{As in Fig 1 for a single power-law description of 
the power spectrum without the QPO from the full lightcurve
(81 d.o.f.)}
\label{fig:l}
\end{figure}

One key problem with this method is that it is {\em linear} while the
real data are known to be non-linear. The physical process which
generates the noise keeps $\sigma_{int}/I$ constant as the source
varies (Uttley \& McHardy 2001). This is quite unlike the simulated
lightcurve properties e.g. for $P(f)\propto \nu^{-\alpha}$ with
$\alpha>1$ (red noise) there is no well defined mean, so no well
defined $\sigma_{i,int}/I_{i}$. Thus the snipped lightcurves made from
this power spectrum have large dispersion in $\sigma_{int}/I$, so a
large range in power spectral normalisation, defined as $\int P(f)df=
(\sigma_{int}/I)^2$. This wide dispersion is unlike that of the real
data, where the noise process is self-similar.  An exponential
transform of the simulated linear lightcurve can recover approximately
the required statistical properties (Uttley, McHardy, \& Vaughan 2005)
but instead we simply choose to put a break in the input power
spectrum at $0.5/T_{data}$. At frequencies below the break the
spectrum is flat ($\alpha$=0) and above the break the spectrum follows
the desired input power spectral shape. The result is a set of
snipped lightcurves of the same length and statistical properties as
the real data, with a well defined mean and dispersion.

\begin{figure}
\begin{center}
\begin{tabular}{l}
 \epsfxsize=8cm \epsfbox{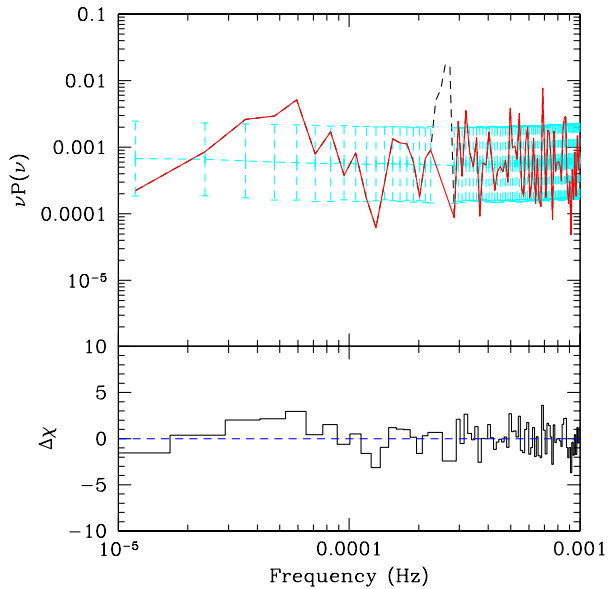}
\end{tabular}
\end{center}
\caption{As in Fig 3 for the best-fit single power-law fit to the power spectrum 
without the QPO from the full lightcurve.}
\label{fig:l}
\end{figure}

\section{The underlying continuum power spectral shape of REJ1034+396}

{\it XMM-Newton} observed REJ1034+396 on 2007-05-31 and 2007-06-01
for $\sim$ 91 ks (observation id. 0506440101, revolution no. 1369).
Details of the data extraction are given in Middleton et al. (2009).
This resulted in 84.3 ks of clean data starting on 2007-05-31 20:10:12
UTC. The QPO is very coherent over 16 periods in a 59ks segment of
this total lightcurve, hereafter called the `in phase' data. The
resulting power spectrum of this `in phase' data is shown in figure 3,
where the QPO is present at $>$99.9\% significance above the red noise
(Gierli{\'n}ksi et al. 2008). The index of the total (i.e. including the
QPO) power spectrum over the frequency range not dominated by
statistical noise (i.e. above 10$^{-3}$Hz) was estimated using the
analytical method of Vaughan (2005) to be -1.35 $\pm$ 0.18 (Gierli{\'n}ski
et al. 2008)

\subsection{Simulations of the in-phase segment}

We first check that our simulations reproduce the analytical result.
Fig 1 shows the $\chi^2$ versus power spectral index for a single
power-law fit to the entire power spectrum above $10^{-3}$~Hz. The
simulations are inherently stochastic so there are fluctuations. We
smooth these by fitting a quadratic which gives a 
minimum at -1.45 $^{+0.20}_{-0.21}$, consistent with the analytical result.

\begin{figure}
\begin{center}
\begin{tabular}{l}
 \epsfxsize=8cm \epsfbox{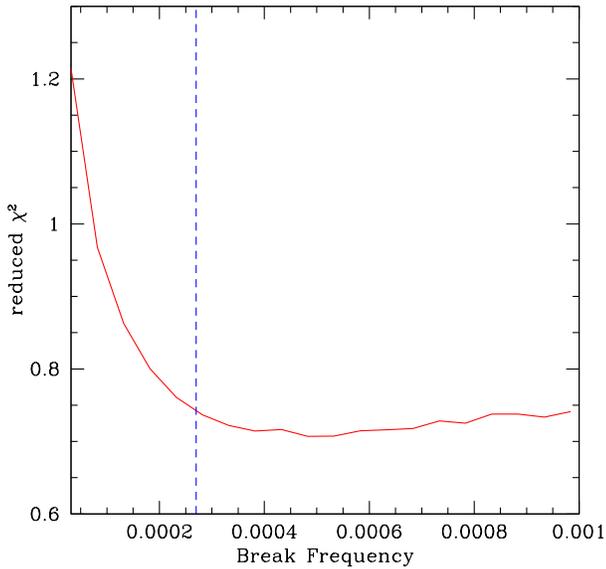}
\end{tabular}
\end{center}
\caption{As in Fig 1 for simulations of a broken power-law 
(indices -1 to -2) fit to the power spectrum without the QPO from the full 
lightcurve. Again, this is not a significantly better fit than the single
power-law shown in Fig 6.}
\label{fig:l}
\end{figure}

However, here we are more interested in the shape of the power
spectrum which underlies the QPO, so we remove the single point at the
QPO frequency, replacing it with the fractional rms of the bin
directly preceding it. We then redo the single power-law simulations
with this reduced ($\sigma_{int}/I)^2$. We show the results from this
in Fig 2, again fitted with a quadratic to smooth over the inherent
stochasticity. This gives -1.35 $^{+0.22}_{-0.21}$ similar to the
previous result as the QPO is fairly central in the fitted section of
the power spectrum, so does not greatly distort the average power
spectral slope. The fit is obviously better in terms of overall
reduced $\chi^2$, and is an acceptable description of the data at
$\sim$ 97\% confidence. This best-fit average power spectrum is shown
together with residuals in Fig 3.

We repeat the analysis using a broken power-law, fixing the slope
below and above the break to -1 and -2 respectively, and then varying
the break frequency. Fig 4 shows that there is a minimum in $\chi^2$
for such a break at a frequency close to the QPO. However, the value
of $\chi^2$ is slightly (though not significantly) higher than that
for a single power-law description of the underlying power spectrum,
so this is not a better description of the power spectrum than a
single power-law. The best-fit, average broken power-law together with
errorbars from statistical uncertainties is shown in figure 5 together
with the residuals to the real power spectrum (minus the QPO) in the
panel below.

\subsection{Full lightcurve}

The longer timescale spanned by the `full' lightcurve extends the
power spectrum down to slightly lower frequencies so may give tighter
constraints on any power spectral break. However, the QPO is now not so
coherent, so is spread over more frequency bins. We simulate the
underlying power spectrum by excluding 5 frequency bins around the QPO
frequency, and replacing them with the preceding bins. Figure 6 shows
the resulting $\chi^2$ distribution for the single power-law, which
has index $-1.08\pm_{-0.18}^{+0.18}$. The overall reduced
$\chi^2\approx 1$, showing that this is a good description of the
data. Figure 7 shows the best-fit average single power-law fit to the
real power spectrum. Error bars come from statistical uncertainties
and, as before, the residuals to the fit are shown in the panel below.

We again test for the presence of a break in the underlying red noise
(excluding the QPO) by simulating a broken power-law with indices fixed
at -1 and -2. The resulting chi-squared plot (figure 8) again shows
that such a spectrum with break frequency slightly above the QPO is
consistent with the data, but is not required by it. As before, the
best-fit broken power-law is plotted in figure 9.

\begin{figure}
\begin{center}
\begin{tabular}{l}
 \epsfxsize=8cm \epsfbox{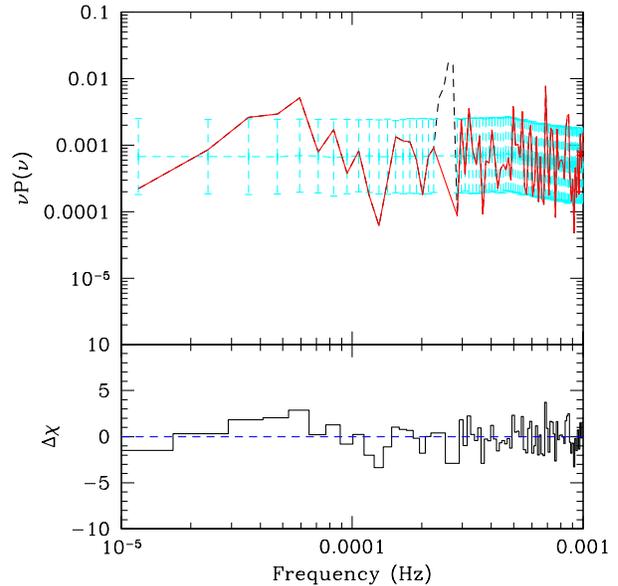}
\end{tabular}
\end{center}
\caption{As in Fig 3 for the best fit broken power-law (indices -1 to -2) fit to the
power spectrum without the QPO from the full lightcurve.}
\label{fig:l}
\end{figure}

\begin{figure*}
\begin{center}
\begin{tabular}{l}
 \epsfxsize=16cm \epsfbox{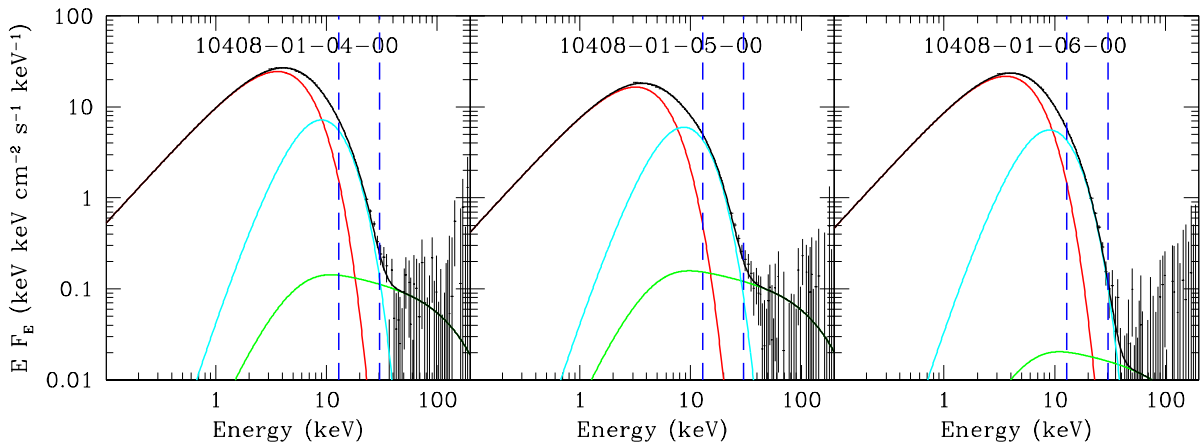}
\end{tabular}
\end{center}
\caption{Unabsorbed PCA and HEXTE X-ray spectra of GRS~1915+105 in the
three observations showing the 67Hz QPO at $>$3$\sigma$. The total
model (black) is made of components from 
a disc blackbody (red), low temperature, optically thick thermal Comptonisation (cyan), 
and thermal Comptonisation with a hotter plasma
(green). The vertical
dashed lines indicate the 13-30~keV energy band used to create the power spectrum 
shown in Fig 12b.} 
\label{fig:l}
\end{figure*}

\subsection{Summary of constraints from the power spectral shape}

The underlying power spectral shape is consistent with a single
power-law with index -1.3. This is inconsistent with the $\sim
\nu^{-2}$ shape seen at high frequencies in the sub-Eddington states
of Cyg X-1 (Revnivtsev et al. 2000), but may be consistent with the
(poorly constrained) noise power shape underlying the HFQPO or 67~Hz
QPO. While this supports our tentative identification of the QPO with
the 67~Hz feature, the power spectral shape is also consistent with
that expected underneath the LFQPO, i.e. with a power spectrum which
breaks from $\nu^{-1}$ to $\nu^{-2}$ at a frequency just above the
LFQPO frequency. However, this can be ruled out by the power spectral
normalisation, which is an order of magnitude below the flat-top
normalisation of $\nu P(\nu) \sim 0.01$ seen ubiquitously under the
LFQPO.  Thus the power spectrum normalisation precludes an
identification with the LFQPO. This supports the conclusions from the
source spectrum and luminosity, which favour a mildly super-Eddington
accretion flow, whereas the LFQPOs are characteristic of sub-Eddington
states.

Instead, the power spectral normalisation is similar to that seen from
the HFQPOs and the 67~Hz QPO from GRS~1915+105.  However, the HFQPOs
are again seen from sub-Eddington states, so the 67~Hz QPO from
GRS1915+105 is our most likely candidate (see Section 2).  Thus we
explore the power spectral correspondence in more detail below.

\section{Energy spectra}

Bandpass effects are very important, as different components have
different variability, and this can substantially affect the derived
normalisation of the power (Churazov et al. 2001; Done \&
Gierli{\'n}ski 2005).  Hence we now look in detail at the energy
spectra for GRS~1915+105 at the time when it shows the 67~Hz QPO, our
most likely candidate for the AGN QPO, and compare components and
bandpass with REJ1034+396.

The data from GRS~1915+105 which show the 67~Hz QPO are the {\it RXTE}
observations on April the 20th (OBSID: 10408-01-04-00), April the 29th
(OBSID: 10408-01-05-00) and May 5th (OBSID: 10408-01-06-00) 1996
(Morgan, Remillard \& Greiner 1997).  The energy spectra of these
observations are shown in Fig 10, fit with a disc black-body component
({\sc diskbb}) and a low-temperature, optically thick thermal Compton
component ({\sc comptt}).  This is accompanied by a tail to higher
energies, which we model as high temperature thermal Comptonisation,
with power-law index fixed at 2.2 and electron temperature at
100~keV. The tail is probably much more complex (see e.g
Zdziarski et al. 2005), but this gives an adequate description of the
{\it RXTE} data.  We show the data with its corresponding best fit model,
corrected for absorption (fixed to the variable abundance model used
in Done, Wardzinski \& Gierli{\'n}ski 2004), superimposed on the data in
Fig 10.

Fig 11 shows the spectrum of REJ1034+396 fit by the same three
component model. This has a ratio of inferred (bolometric) disc flux
to Comptonised flux of 3.0, and a ratio of disc to power-law flux of
19, similar to that of GRS~1915+105 (4.3/38, 3.0/24 and 4.3/33 for the
04, 05 and 06 ObsIDs respectively). However, the ratio of electron
temperature to disc temperature is rather higher for REJ1034+396 than
for GRS~1915+105 (6.6 versus 1.6, 1.8 and 1.7), making the
approximately power-law section of the `soft excess' rather than Wien
shape of this component. There are other energy spectra in
GRS~1915+105 which match this aspect more closely (Middleton et
al. 2009), but these data do not show a significant high frequency
QPO.

\begin{figure}
\begin{center}
\begin{tabular}{l}
 \epsfxsize=8cm \epsfbox{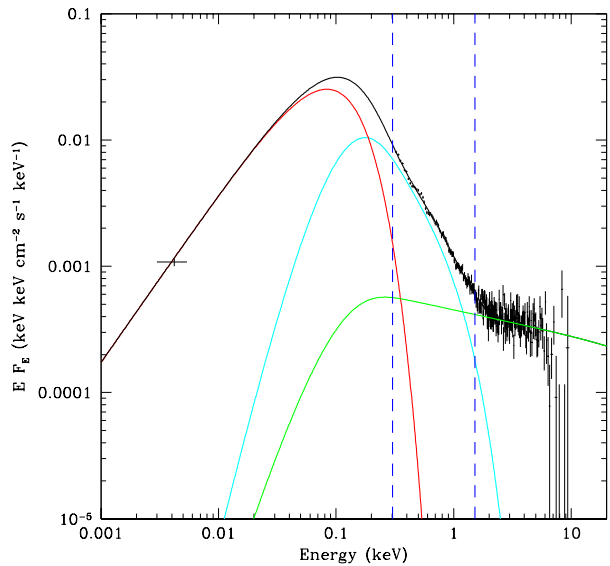}
\end{tabular}
\end{center}
\caption{Unabsorbed Optical/UV/X-ray SED of REJ1034+396 showing the
deconvolved disc (red), low-temperature Compton component (cyan) and
Comptonisation from a higher temperature plasma (green). The vertical
dashed lines indicate the energy band which contains a similar amount
of the spectral components (dominated by the low temperature Comptonisation)
as in the 13-30~keV energy band in 
GRS~1915+105 (Fig 10).}
\label{fig:l}
\end{figure}

\begin{figure*}
\begin{center}
\begin{tabular}{l}
 \epsfxsize=16cm \epsfbox{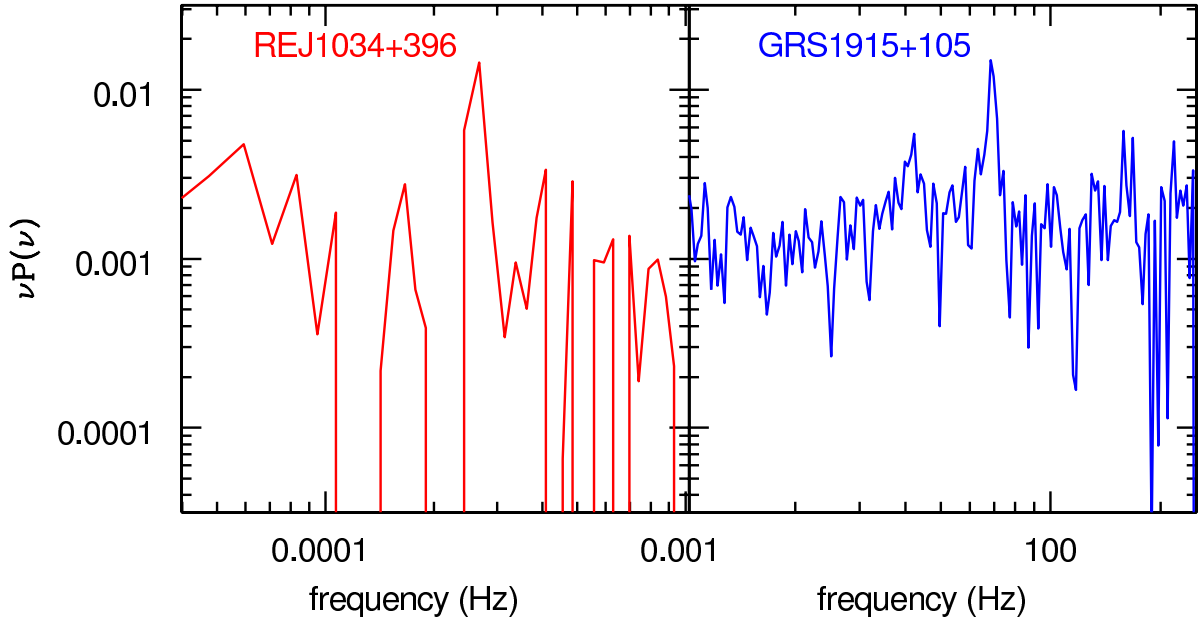}
\end{tabular}
\end{center}
\caption{(a) shows the PDS of REJ1034+396 in the
0.3-1.5keV energy band, while (b) shows the corresponding 
energy bandpass 13-30~keV PDS from GRS~1915+105
(Remillard et al. 2003). Unlike the other power spectra shown in this
paper, the statistical noise has been subtracted from both, so that the
PDS shows only the intrinsic power. 
Both cover 1.5 decades in frequency, and are on the same scale. 
Both normalisation and slope underlying the QPO are very similar, supporting 
the identification of the AGN QPO as analogous to the 67~Hz QPO in GRS~1915+105.}
\label{fig:l}
\end{figure*}

We can now quantify the bandpass effects. In GRS~1915+105, Remillard
et al. (2004) show the 41/67~Hz QPO computed from the lightcurve in
the 13-30~keV bandpass range, rather than the 2--20~keV in Morgan et
al (1997).  We overlay this energy range on the spectra in Fig 10 to
show that this samples mostly the low temperature Comptonisation
component, excluding most of the disc component which dilutes the
variability, as well as most of the higher energy power-law. To be
dominated by the corresponding spectral components in REJ1034+396
requires a bandpass of 0.3-1.5~keV.

Fig 12a shows the resulting power spectrum from 0.3-1.5~keV. Unlike
all the other power spectra, this one has statistical noise
subtracted, so has less power at high frequencies. However, the
intrinsic power is very similar to that of the total lightcurve in
both shape and normalisation.  Fig 12b shows the 13-30~keV power
spectrum of GRS~1915+105 of Remillard et al. (2004), again with
statistical noise subtracted. This is plotted using the same scale in
$\nu P(\nu)$ over 1.5 decades in frequency (10-300Hz in GRS~1915+105,
scaled by the putative mass difference of $3\times 10^6/14$
corresponding to $4.66\times 10^{-5}-1.4\times 10^{-3}$~Hz in
REJ1034+396). While the power spectrum of the AGN has very poor
statistics compared to GRS 1915+105, these data are startlingly similar
in both shape and normalisation for both the QPO and underlying power
spectrum. Much better data would be required in order to constrain the
existence of any 3:2 harmonic corresponding to the 41~Hz QPO in
GRS 1915+105 (it is barely significant in GRS 1915+105!). The slight
excess of power at this frequency in the AGN is not at all
significant.

\section{Conclusion}

The discovery of an AGN QPO allows us to test whether the physics of
the accretion flow scales linearly with black hole mass if we can
identify the corresponding BHB QPO. We use the variability power
spectral shape and normalisation to show that the continuum noise
underlying the AGN QPO is consistent with that seen underneath the
HFQPO seen in BHB and 67~Hz QPO seen in GRS1915+105, but {\em not}
with the LFQPO.

The distinctive shape of the energy spectrum of REJ1034+396 argues
strongly for a mildly super-Eddington accretion flow in this AGN,
supported by recent mass estimates for the black hole (Bian \& Huang
2009). This favours the 67~Hz QPO as the BHB counterpart, as this is
seen in super-Eddington states of GRS 1915+105, whereas the HFQPOs are
seen in sub-Eddington states. The spectra seen from the 67~Hz QPO
observations of GRS 1915+105 are similar to that the REJ1034+396 in
being dominated by low temperature, optically thick Comptonisation of
the disc. This spectral shape may well be the new `ultra-luminous
state' proposed by Gladstone et al. (2009) for the broad curvature
seen in the X-ray spectra of the Ultra-Luminous X-ray
sources. Conversely, the disc is seen without strong distortion by
Comptonisation in the spectra corresponding to the HFQPOs.

The identification of the AGN QPO at $2.7\times 10^{-4}$Hz with the
67~Hz QPO in GRS 1915+105 is consistent with a simple {\em linear} mass
scaling of the QPO frequency, energy spectral components and
variability power spectra from stellar to supermassive black holes
accreting at the same $L/L_{Edd}$.

\section{Acknowledgements}

We would like to thank NASA and ESA for making the data available for
{\it RXTE} and {\it XMM-Newton} via the HEASRC website. This work was carried out
through an STFC PhD grant. We would also like to thank Ron Remillard
at MIT for helpful discussions and providing the data for the power
spectrum of GRS~1915+105 in figure 12.

\label{lastpage}

\label{lastpage}

\end{document}